\documentclass{aastex} 
\usepackage[numberedappendix]{emulateapj5} 
\usepackage{apjfonts}
\usepackage{natbib}

\journalinfo{To appear in The Letters of the Astrophysical.  J.}

\slugcomment{VERSION Accepted June 20, 2001}
\shortauthors{Muench et al.} 
\shorttitle{Brown Dwarfs in the Trapezium}

\newcommand{\solarmass}{\ensuremath{ \mbox{M}_{\sun} }}
\newcommand{\av}{\ensuremath{ \mbox{A}_{V} }}
\newcommand{\Ks}{\ensuremath{ \mbox{K}_{\mbox{s}} }}

\begin{document}

\title{Evidence for Circumstellar Disks around Young Brown Dwarfs \\
in the Trapezium Cluster}

\author{ August A.  Muench \altaffilmark{1,2,3},
Jo\~{a}o Alves \altaffilmark{4,5},
Charles J.  Lada \altaffilmark{2,5},
and Elizabeth A.  Lada \altaffilmark{1} }

\altaffiltext{1}{Department of Astronomy, University of Florida,
  Gainesville, FL 32611; muench@astro.ufl.edu, lada@astro.ufl.edu}
\altaffiltext{2}{Harvard-Smithsonian Center for Astrophysics, Cambridge,
  MA 02138; clada@cfa.harvard.edu}
\altaffiltext{3}{Smithsonian Predoctoral Fellow}
\altaffiltext{4}{European Southern Observatory,
  Karl-Schwartzschild-Strasse 2, 85748 Garching, Germany; jalves@eso.org}
\altaffiltext{5}{Visiting Astronomer, European Southern Observatory}

\begin{abstract}
 We report the results of deep infrared observations of 
brown dwarf candidates in the Trapezium cluster in Orion.  Analysis of the
JHK color-color diagram indicates that a large fraction
($\sim\,65\%\,\pm\,15\%$) of the observed sources exhibit infrared excess
emission.  This suggests the extreme youth of these objects and in turn,
provides strong independent confirmation of the existence of a large
population of substellar objects in the cluster.  Moreover, this suggests
that the majority of these substellar objects are presently surrounded by
circumstellar disks similar to the situation for the stellar population of
the cluster.  This evidence for a high initial disk frequency ($>$ 50 \%)
around cluster members of all masses combined with the smooth continuity of
the cluster's initial mass function across the hydrogen burning limit
suggests that a single physical mechanism is likely responsible for
producing the entire cluster mass spectrum down to near the deuterium
burning limit.  The results may also indicate that even substellar objects
are capable of forming with systems of planetary companions.
\end{abstract}


\section{Introduction} 
\label{sec:intro}

Among the most fundamental issues raised by the existence of brown dwarfs
is the question of their origin and genetic relationship to planets and
stars.  Are brown dwarfs giant planets or small, failed stars, or,
something else altogether different?  The critical test needed to resolve
this question is to determine whether brown dwarfs primarily form within
circumstellar disks as companions to stars, similar to planets, or from
their own individual cloud cores or fragments, like stars.  To date, the
most important observations bearing on this question have been:  1) the
observed lack of close brown dwarf companions found in radial velocity
surveys of nearby field stars \citep[the so-called brown dwarf desert,
e.g.,][]{marcy98} and 2) the existence of free floating brown dwarfs in
young clusters \citep[e.g.,][]{bouv98}.  Both facts would appear to
implicate a stellar (non-planet like) origin for these objects, i.e.,
formation from independent, contracting fragments of the parental molecular
cloud.  However, our understanding of the origin of substellar objects is
far from complete.  For example, an alternative formation scenario
has been recently proposed by \citet{rc2001} who suggest that most
freely floating brown dwarfs did
not form from their own protostellar fragments, but instead were initially
formed as companions to other protostars and then were dynamically
ejected via 3 body encounters before they could grow into stellar mass objects.

The most direct way to address the question of the origin and nature of
brown dwarfs is to investigate the properties of extremely young substellar
objects in regions of active star and planet formation.  For example,
finding young brown dwarfs surrounded by their own circumstellar accretion
disks would likely implicate a stellar-like formation mechanism (from
individual cloud fragments) and place strong constraints for the
theoretical models of their origin \citep[e.g.,][]{rc2001}.  Moreover, such
a finding would raise the interesting question of whether planetary
companions can form around such objects.

Recently, \citet{lada2000} used near-infrared ($1\,-\,3\,\micron$) 
color-color diagrams to show that a large fraction
($\sim\,80\,-\,85\,\%$) of the stars in the young Trapezium Cluster display
thermal infrared excess indicative of circumstellar disks.  Further, they
found that the fraction of stars with disks remained high with decreasing
mass to near the hydrogen burning limit.  Below this limit, their
observations became incomplete.  Does the incidence of circumstellar disks
also continuously extend across the hydrogen burning limit to substellar
mass objects?

Deeper infrared observations reveal a substantial population of faint
sources which could be free floating substellar objects in this cluster
\citep{mcc95,mll2000,lr2000,hc2000,luh2000}.  However, the identifications of
these sources as substellar objects are not secure because observations of
nearby unextincted control fields reveal significant numbers of field stars
in the corresponding brightness range \citep*[][]{mlla01}, suggesting that
field star contamination could be a severe problem, especially for the
faintest candidates.  Reasonable attempts to account for the effects of the
screen of extinction provided by the molecular cloud behind the Trapezium
do suggest that the vast majority of the brown dwarf candidates are not
reddened field stars \citep{hc2000, mlla01}.  However, independent
confirmation of membership is clearly important and could be provided by
indications of extreme youth, such as the presence of infrared excess and
dusty disks surrounding these objects.

In this letter, we present an observational analysis of a deeper and more
complete set of near-infrared observations for the candidate brown dwarf
population in the Trapezium Cluster.  We find a relatively large fraction
of the candidates exhibit infrared excess indicative of circumstellar
disks.  This confirms both their membership in the cluster and their status
as substellar objects and perhaps suggests an origin for them that is more
stellar-like than planetary-like.

\section{Observations}
\label{sec:data}

We obtained deep JH\Ks~images of the central $5\arcmin\:\times\:5\arcmin$
region of the Trapezium Cluster during 1 hour on 14 March 2000 using the
SOFI infrared camera on the ESO 3.5~m New Technology Telescope in La Silla,
Chile.  The NTT telescope uses an active optics platform to achieve ambient
seeing and high image quality and the SOFI camera employs a large format
$1024\:\times\:1024$ pixel Hawaii HgCdTe array.  We configured SOFI to have
a $\:4\farcm95\:\times\:4\farcm95\:$ field of view with a plate scale of
0\farcs29 /pixel.  Seeing conditions were superb ($\sim\,0.55\arcsec$) and
our 90\% completeness limit is estimated at $\Ks\,\sim\,17.5$.  These NTT
observations were obtained as part of larger program to catalog the
infrared JHKL magnitudes of sources over the entire mass spectrum in the
Trapezium Cluster.  The brighter portion of this catalog was presented in
\citet{lada2000}, and only the NTT observations are discussed in this paper.
A complete description and analysis of these observations is presented 
in the forthcoming \citet[ hereafter MLLA2001 ]{mlla01}.

\section{Results and Analysis}
\label{sec:results}

In figure \ref{fig:trap_faint_locus_cmag} we construct the infrared
color-magnitude diagrams for those NTT Trapezium sources which were
simultaneously detected at JH\Ks~ wavelengths.  In these diagrams, we
compare the locations of these sources to the location of the theoretical
isochrone from the \citet[ hereafter BCAH98]{bcah98} non-grey evolutionary
models at the assumed mean age (1 Myrs) and distance (400 pc) of this
cluster.  The BCAH98 theoretical isochrone closely follows the near-IR
colors of the Trapezium sources, forming an excellent left-hand boundary to
the source distribution in this color-magnitude space.  The Trapezium
sources are reddened away from this boundary with implied extinctions of
$\av\,\sim\,1-35\,\mbox{mag}$.  

We identified candidate brown dwarfs in the Trapezium Cluster by comparing
the infrared luminosities of detected sources to those predicted by the
theoretical evolutionary models.  We selected all the NTT sources in the
J-H/H diagram (figure \ref{fig:trap_faint_locus_cmag}a) fainter than the
predicted luminosity of the hydrogen burning
limit (hereafter HBL; 0.08 \solarmass)\footnote{The predicted colors and
magnitudes of the hydrogen burning limit for this distance/age combination
are essentially identical to those for a younger assumed age (0.4 Myrs) but
at a larger distance (470pc).}  but brighter than the luminosity of an 0.02
\solarmass~object.  This lower limit was chosen because the current
theoretical evolutionary models do not extend much below this mass, and
because we wish to exclude cooler, lower mass objects whose intrinsic
colors are not well constrained.  Between these two mass/luminosity limits,
we identified 112 candidate brown dwarfs in the J-H/H diagram.  
We also indicate in Figure 1 the locations of 10 Trapezium
Cluster members with spectral types
equal to or later than M6 in \citet[][]{lah97}.  The spectral type M6 is an
important boundary because recent spectroscopic studies have suggested that
it represents the hydrogen burning limit in very young
($\tau\,\lesssim\,10\,$Myrs) clusters \citep{luh99}.  In figure
\ref{fig:trap_faint_locus_cmag}a, these late type sources are on average 1
magnitude brighter than our adopted hydrogen
burning limit.  The faintness of our IR selected brown dwarfs relative to
these late type sources confirms that we are likely selecting sources below
the HBL.

We refine our selection of brown dwarf candidates by plotting the J-H/H
candidates in the H-\Ks/\Ks~color-magnitude diagram in figure
\ref{fig:trap_faint_locus_cmag}b. 
In this diagram a number of candidates are brighter and redder than the
hydrogen burning limit.  We retain these as likely brown
dwarf candidates because they have photometric errors which are much too
small to have scattered them to this location, because they are fainter than
most of the M6+ dwarfs, and because excess $2\,\micron$ flux from
circumstellar disks could act to brighten and redden such sources 
out of the brown dwarf regime in the H-\Ks/\Ks~color-magnitude diagram.
A few very faint candidates scatter below the 0.02 \solarmass~limit 
in the H-\Ks/\Ks~diagram, and we exclude these sources from our final sample.

In figure \ref{fig:trap_faint_locus_jhk}, we plot the H-\Ks/J-H color-color
diagram for the 109 candidate brown dwarfs.  We also plot for comparison
the loci of colors for giants and for main-sequence dwarfs from
\citet{bb88}.  We extended the loci of M dwarf colors in figure
\ref{fig:trap_faint_locus_jhk} from M6 to M9 using the empirical brown
dwarf colors compiled in \citet{kirk2000}.  The predicted effective
temperatures of 1 Myr brown dwarfs from the BCAH98 evolutionary models are
quite warm, e.g. $\mbox{T}_{eff}\,\ge\,2500\mbox{K}$ for masses
greater than our 0.02 \solarmass~limit.  Therefore, we
expect that the intrinsic infrared colors of such young brown dwarfs
are no redder than those of M9 dwarfs \citep[J-H = 0.72;
H-\Ks~= 0.46][]{kirk2000} which agree well with the H-K colors predicted by
current model atmospheres of low surface gravity, 2600K sources
\citep{allard2001}\footnote{and also
\url{ftp://ftp.ens-lyon.fr/pub/users/CRAL/fallard/}}.

We find $65\%\,\pm\,15\%$ (71/109) of our candidates fall to the right of
the reddening band for M dwarfs and into the infrared excess region of the
color-color diagram.  
We further determine that 54\% of the candidates have an infrared excess 
that is greater than their $1\sigma$ photometric uncertainties in color.  
In addition to normal photometric
uncertainties the measured colors of these sources could be corrupted by the
presence of the strong nebular background, and we performed an extensive set
of artificial star photometry experiments to test this possibility. 
We found that nebular contamination can introduce some additional scatter 
to a star's measured J-H color and this can explain in part the 
J-H colors of $\sim\,25\%$ of the excess sources which are bluer than
expected for late type sources ($\mbox{J-H}\,<\,0.6$).
However, blueward J-H scatter can produce a false excess fraction
($\sim\,10-20\%$) only for the the faintest artificial stars,
i.e., H = K = $\sim\,16$ mag.
Further, we find that such nebular contamination never produces as large a 
dispersion of the H-\Ks~colors as found in our observations of the
candidate brown dwarfs, and we conclude that the observed infrared excesses
are an intrinsic property of these objects.

\section{Discussion and Conclusions}
\label{sec:discuss}

From analysis of their near-infrared colors, we find that $\sim\,50\%$ of
the candidate brown dwarfs in the Trapezium cluster display significant
near-infrared excess.  This is similar to the behavior of the
stellar population of this
cluster and suggests the extreme youth of these low luminosity sources.
This, in turn, provides independent confirmation of their membership in the
cluster and their nature as bona fide substellar objects.  
As is the case for the more massive stellar members, the most
likely explanation for the observed near-infrared excesses around 
the brown dwarfs in this cluster is the presence of circumstellar disks.
Strong, independent support for the disk interpretation derives from
the fact that we find  21 of the candidate brown dwarfs to be
spatially coincident with  optically identified HST ``proplyds''
\citep{bally00,odw96} which are known to be photo-evaporating
circumstellar disks.
We note that the proplyd brown dwarfs display
a JHK excess fraction of 71\%, while the brown dwarf candidates
unassociated with known proplyds have a slightly lower excess
fraction of 63\%.
The proplyd brown dwarfs also display bluer J-H colors than
the remaining brown dwarf candidates and account for half the
excess sources with J-H color $<\,0.6$.
Despite their relatively blue J-H colors, the proplyd nature
of these sources affirms the hypothesis that the observed JHK 
infrared excess is intrinsic and a signature of the presence of a
circumstellar disk.

The hypothesis that the observed near-IR excess is caused by circumstellar
disks is further supported by observations of brown dwarf candidates in other
clusters.  Late-type brown dwarf candidates in the $\rho$ Ophiuchi cluster
were identified by their water vapor absorption features and display
evidence for veiling in their infrared spectra as well as evidence for
infrared excesses in their H-K/J-H color-color diagrams \citep{wilk99,
cush2000}.  ISO ($6.7\micron$) observations reveal 4 brown dwarf candidates
with mid-infrared excesses in Chamaeleon \citep{cnk2000}.  \citet{luh99}
identified 7 brown dwarf candidates in the IC 348 cluster which, after
de-reddening, fall to the right of the main-sequence reddening band but
below the cTTS locus similar to the locus of brown dwarfs identified here.
Luhman also identified strong $\mbox{H}\alpha$ emission
($\mbox{W}[\mbox{H}\alpha]\,>\,10\,\mbox{\AA}$) in a number of these sources and
suggested that these are not simply passive circumstellar disks, but that
these brown dwarfs are undergoing accretion.  Finally, powerful evidence
for accretion disks around very young brown dwarfs was found by
\citet{muz00} who identified an asymmetric $\mbox{H}\alpha$ emission line
profile for the M6 PMS object V410 Anon 13 in Taurus and successfully used
magnetospheric accretion models to show that this brown dwarf candidate was
indeed accreting but at a rate much lower than has been found in higher
mass stars.

Compared to these other studies, our sample of Trapezium Cluster brown
dwarfs is the first population that is sufficiently large to statistically
estimate the frequency of substellar objects born with circumstellar disks.
Indeed, there are now more brown dwarfs identified in the Trapezium cluster
than are presently known in all other star forming regions combined.
However, our estimate of the disk frequency from the JHK diagram could
underestimate the true disk frequency for a number of reasons. 
First, JHK observations trace the innermost regions of disks
and the particular disk geometry (inclination, presence of inner disk holes,
etc.) can act to reduce the efficiency of detecting disks from JHK photometry,
especially for late type sources \citep{hil98,la92}.  For example,
the $\sim\,50\%$ excess fraction we find for brown dwarfs is nearly
identical to the excess fraction found in the JHK diagram of
\citet{lada2000} for objects in the cluster which are above the hydrogen
burning limit.  However, by employing $3 \micron$ photometry, these authors
found a much higher, $\sim\,85\%$, disk frequency even for the faintest
members they detected.  Indeed of the 10 M6+ sources shown in figure
\ref{fig:trap_faint_locus_cmag}, 9 were detected at L band and although
only one of these sources clearly displays H-K excess, 8 display K-L
excess.  Second, as a result of selecting candidate brown dwarfs at all
reddenings we may have included reddened background field stars in the
sample which would act to decrease the derived disk fraction.  When we
select candidates at low reddenings ($\av\,\le\,5$ relative to the
isochrone in the figure \ref{fig:trap_faint_locus_cmag}a) to exclude
background field stars, we find 77\% (57/74) of this sample display
infrared excess.  Further, this sample is an extinction
limited sample which is complete at all masses in our selected range
and therefore is likely representative of the population as a whole.

We conclude from our current study and from the findings of
\citet{lada2000} that circumstellar disks are present around a high
fraction of Trapezium Cluster members {\em across the entire mass
spectrum}.  This implies that brown dwarfs and higher mass stars form via a
similar mechanism, e.g., from individual contracting fragments of the
parental molecular cloud which, via conservation of angular momentum, form
a central star  accompanied by a circumstellar disk \citep{sal87}.
\citet{low76} showed that within the conditions of molecular clouds, the
minimum Jeans mass for a cloud fragment could be as small as 0.007
\solarmass, well below the mass necessary to create the Trapezium brown
dwarfs. The free-floating nature of these brown dwarfs
rules out their formation as companions in a circumstellar disk unless
the sources were ejected. But models of dynamical ejection of these
objects from hierarchical systems predict
that any circumstellar material will be
disrupted during the ejection process and that circumstellar disks
should be rare among ejected brown dwarfs 
\citep[e.g., ][]{rc2001}.
Thus our results seem to implicate a formation mechanism for brown dwarfs
in which such objects are formed with circumstellar disks from individual 
proto-{\it sub}stellar cores.
Consequently, even sub-stellar objects may be capable of forming with 
systems of planetary companions.

Confirmation of our hypothesis that a substantial fraction of the brown
dwarfs in the Trapezium are surrounded by circumstellar disks
requires additional data.  Deep $3 \micron$ ground-based observations such
as those used by \citet{lada2000} are necessary to permit a more accurate
measurement of the excess fraction for the brown dwarf population.
Longer wavelength infrared observations, such as those that will be
possible with SIRTF, would enable the construction of more complete SEDs
for these sources which could then be compared directly to theoretical disk
predictions.  Estimates of the masses of the disks would have
interesting implications for the possibility of forming planetary
companions around brown dwarfs.  Finally, high resolution spectra of these
objects would enable searches for accretion indicators, such as H$\alpha$
emission, veiling, etc., which could yield accretion rates and information
about the growth and early evolution of these interesting objects.

\acknowledgments We thank John Stauffer for his suggestion to examine the
colors predicted by non-grey model atmospheres and Kevin Luhman for
comments and suggestions on an earlier version of this work.  AAM was
supported by a Smithsonian Predoctoral Fellowship and by the NASA Graduate
Student Research Program.  EAL and AAM acknowledge support from a Research
Corporation Innovation Award and Presidential Early Career Award for
Scientists and Engineers (NSF AST 9733367) to the University of Florida.

\newpage

\clearpage 
\thispagestyle{myheadings} 
\markright{Figure 1 }
\setcounter{page}{1}

\includegraphics[width=4.5in,angle=90]{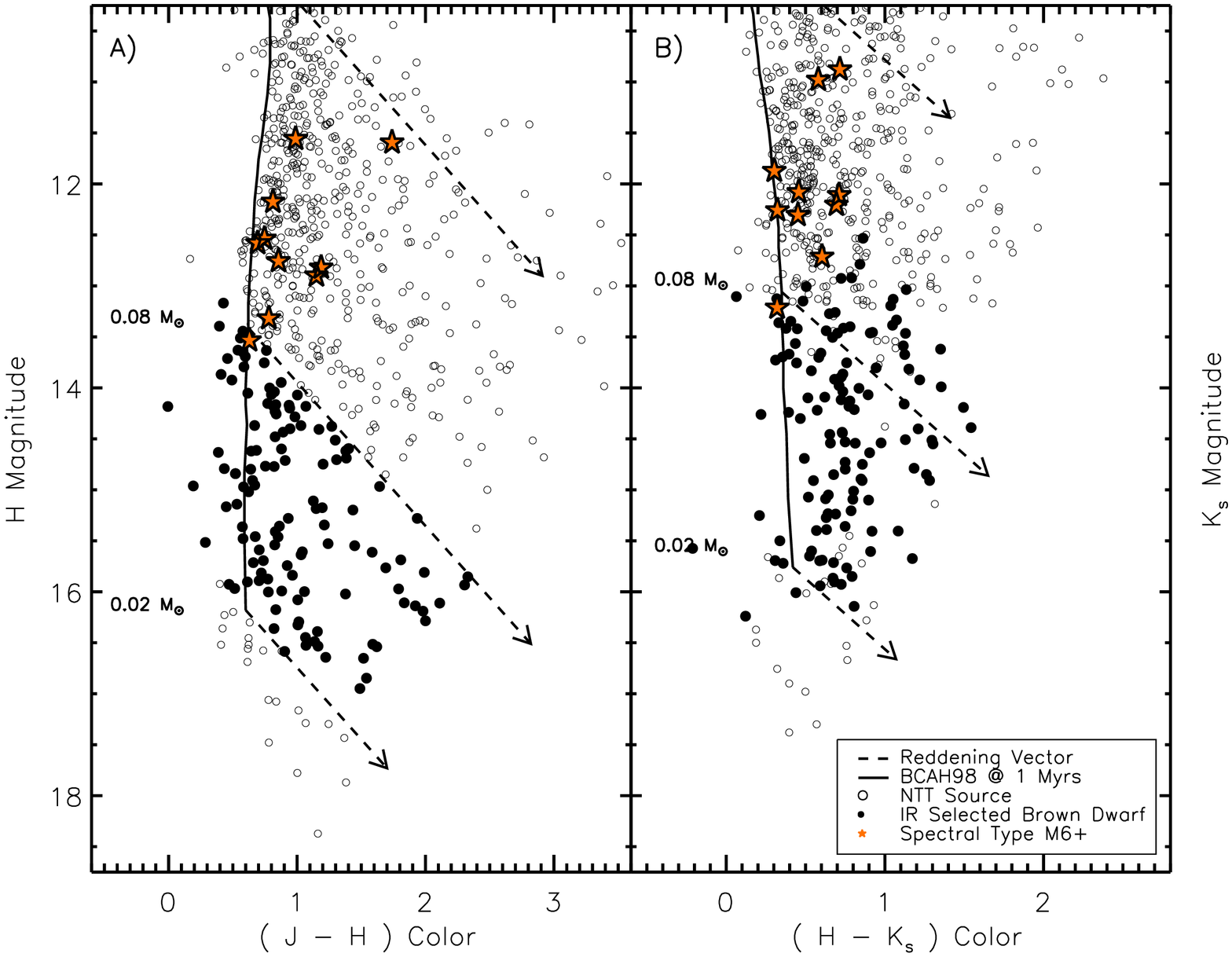}

\figcaption[fig.trap_faint_locus_a.eps]{
Candidate Brown Dwarfs from the Trapezium Cluster Infrared Color-Magnitude 
Diagrams.  Only sources having NTT observations and JH and \Ks~magnitudes 
are shown.  Trapezium sources are compared to the location 
of the 1 Myr (at 400pc) isochrone from the BCAH98 model atmospheres.  
Candidate brown dwarfs (filled circles) were selected by their H band 
luminosities (and colors) and are marked in both color-magnitude diagrams.  
Reddening vectors for 1, 0.08 and 0.02 \solarmass~objects are drawn at 
visual extinctions of 20, 20 and 10 magnitudes, respectively.  Stars with 
spectral types $\geq\,\mbox{M6}$ are identified as filled stars.  
a) J-H/H color-magnitude diagram.  
b) H-\Ks/\Ks~color-magnitude diagram.  
\label{fig:trap_faint_locus_cmag}
}

\clearpage 
\newpage 
\thispagestyle{myheadings} 
\markright{Figure 2 }
\setcounter{page}{2}

\includegraphics[width=5.5in,angle=90]{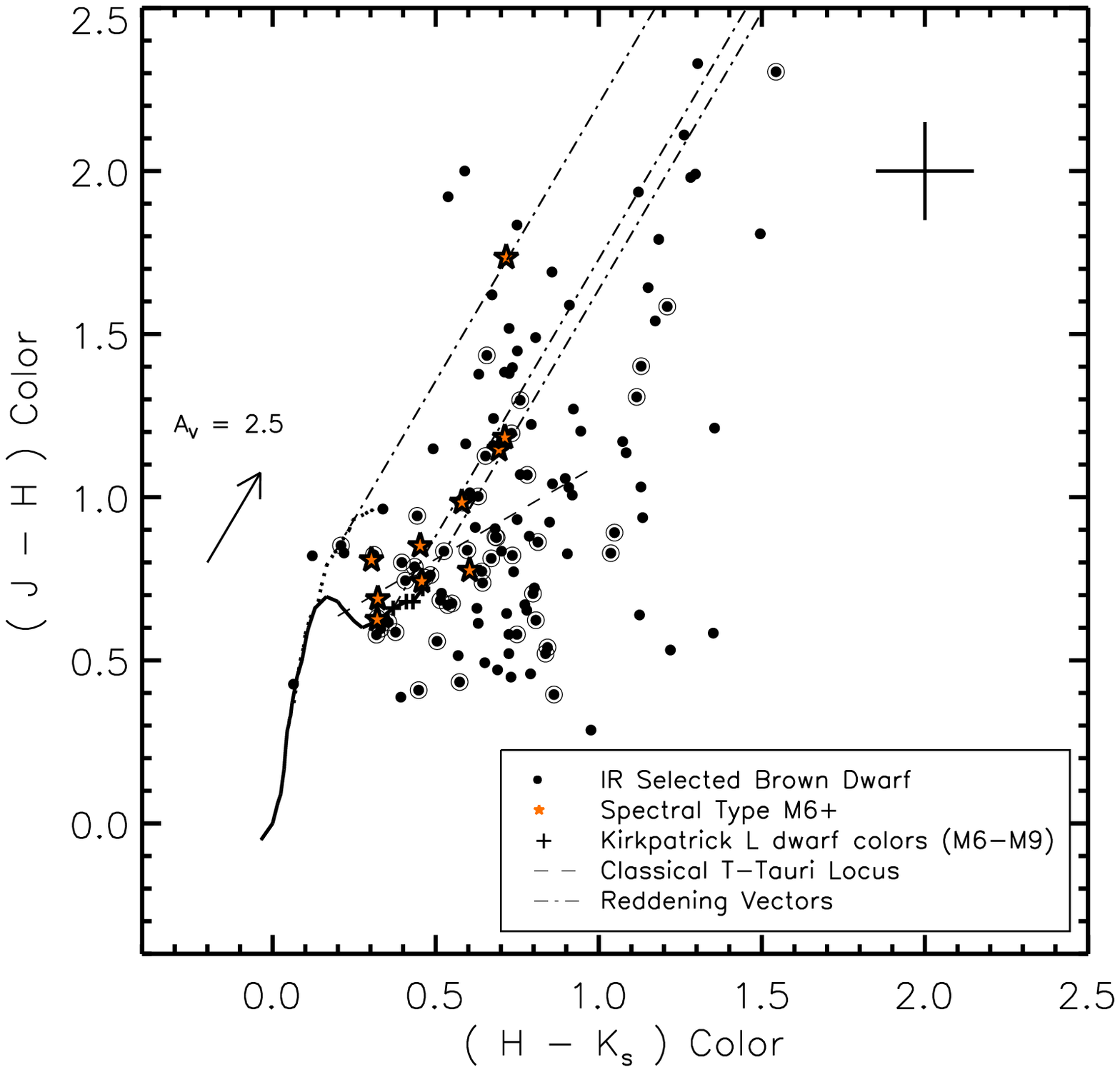}

\figcaption[fig.trap_faint_locus_b.eps]{
H-\Ks/J-H color-color diagram for the  109 sources with NTT JH\Ks~magnitudes 
and which fall into brown dwarf regime of Fig 1a.  The candidate brown dwarfs 
are compared to the intrinsic colors of giants and A0-M6 dwarfs from 
\citet{bb88}, the late M (M6 - M9) color sequence from \citet{kirk2000} and 
the Classical T-Tauri locus from \citet{mch97}.  Appropriate reddening vectors
\citep{coh81} are drawn for giants, for M6 stars and for M9 stars.  Colors of 
Trapezium sources with very late (M6+) spectral types are shown as stars.  
Circled sources have color errors of less than 10\% and the size of 15\% 
uncertainties in color are illustrated at the upper right. 
\label{fig:trap_faint_locus_jhk}
}

\end{document}